\documentclass[twocolumn]{aastex62}

\usepackage{amsmath}
\usepackage{amssymb}
\usepackage{rotating}

\newcommand\tE{t_{\rm E}}
\newcommand\thetaE{\theta_{\rm E}}

\newcommand\murel{\mu_{\rm rel}}
\newcommand\pirel{\pi_{\rm rel}}
\newcommand{\ra}[4]{${#1}^{\rm h}{#2}^{\rm m}{#3}\fs{#4}$}
\newcommand{\dec}[4]{${#1}\arcdeg{#2}\arcmin{#3}\farcs{#4}$}

\newcommand\jkas{J. Korean Astron. Soc.}

\received{January 1, 2018}
\revised{January 1, 2018}
\accepted{January 1, 2018}

\shorttitle{A terrestrial-mass planet detected in the shortest-timescale microlensing event}
\shortauthors{Mr\'oz et al.}

\begin{document}

\title{A terrestrial-mass rogue planet candidate detected in the shortest-timescale microlensing event}

\correspondingauthor{Przemek Mr\'oz}
\email{pmroz@astro.caltech.edu}

\author[0000-0001-7016-1692]{Przemek Mr\'oz}
\affil{Division of Physics, Mathematics, and Astronomy, California Institute of Technology, Pasadena, CA 91125, USA}
\affil{Astronomical Observatory, University of Warsaw, Al. Ujazdowskie 4, 00-478 Warszawa, Poland}

\author[0000-0002-9245-6368]{Rados\l{}aw Poleski}
\affil{Astronomical Observatory, University of Warsaw, Al. Ujazdowskie 4, 00-478 Warszawa, Poland}

\author{Andrew Gould}
\affil{Max Planck Institute for Astronomy, K\"{o}nigstuhl 17, D-69117 Heidelberg, Germany}
\affil{Department of Astronomy, Ohio State University, 140 W. 18th Ave., Columbus, OH 43210, USA}

\author[0000-0001-5207-5619]{Andrzej Udalski}
\affil{Astronomical Observatory, University of Warsaw, Al. Ujazdowskie 4, 00-478 Warszawa, Poland}

\author{Takahiro Sumi}
\affil{Department of Earth and Space Science, Graduate School of Science, Osaka University, Toyonaka, Osaka 560-0043, Japan}

\nocollaboration{} 

\author[0000-0002-0548-8995]{Micha\l{} K. Szyma\'nski}
\affil{Astronomical Observatory, University of Warsaw, Al. Ujazdowskie 4, 00-478 Warszawa, Poland}

\author[0000-0002-7777-0842]{Igor Soszy\'nski}
\affil{Astronomical Observatory, University of Warsaw, Al. Ujazdowskie 4, 00-478 Warszawa, Poland}

\author[0000-0002-2339-5899]{Pawe\l{} Pietrukowicz}
\affil{Astronomical Observatory, University of Warsaw, Al. Ujazdowskie 4, 00-478 Warszawa, Poland}

\author[0000-0003-4084-880X]{Szymon Koz\l{}owski}
\affil{Astronomical Observatory, University of Warsaw, Al. Ujazdowskie 4, 00-478 Warszawa, Poland}

\author[0000-0002-2335-1730]{Jan Skowron}
\affil{Astronomical Observatory, University of Warsaw, Al. Ujazdowskie 4, 00-478 Warszawa, Poland}

\author[0000-0001-6364-408X]{Krzysztof Ulaczyk}
\affil{Department of Physics, University of Warwick, Coventry CV4 7 AL, UK}
\affil{Astronomical Observatory, University of Warsaw, Al. Ujazdowskie 4, 00-478 Warszawa, Poland}

\collaboration{(OGLE Collaboration)}
\noaffiliation

\author{Michael D. Albrow}
\affil{University of Canterbury, Department of Physics and Astronomy,
Private Bag 4800, Christchurch 8020, New Zealand}

\author{Sun-Ju Chung}
\affil{Korea Astronomy and Space Science Institute, Daejon 34055,
Republic of Korea}
\affil{University of Science and Technology, Korea (UST) Gajeong-ro,
Yuseong-gu,  Daejeon 34113, Republic of Korea}

\author{Cheongho Han}
\affil{Department of Physics, Chungbuk National University, 
Cheongju 28644, Republic of Korea}

\author{Kyu-Ha Hwang}
\affil{Korea Astronomy and Space Science Institute, Daejon 34055,
Republic of Korea}

\author{Youn Kil Jung}
\affil{Korea Astronomy and Space Science Institute, Daejon 34055,
Republic of Korea}

\author{Hyoun-Woo Kim}
\affil{Korea Astronomy and Space Science Institute, Daejon 34055,
Republic of Korea}

\author{Yoon-Hyun Ryu}
\affil{Korea Astronomy and Space Science Institute, Daejon 34055,
Republic of Korea}

\author{In-Gu Shin}
\affil{Korea Astronomy and Space Science Institute, Daejon 34055,
Republic of Korea}

\author{Yossi Shvartzvald}
\affil{Department of Particle Physics and Astrophysics, Weizmann 
Institute of Science, Rehovot 76100, Israel}

\author{Jennifer C. Yee}
\affil{Center for Astrophysics $|$ Harvard \& Smithsonian, 60 Garden St., 
Cambridge, MA 02138, USA}

\author{Weicheng Zang}
\affil{Department of Astronomy and Tsinghua Centre for Astrophysics, 
Tsinghua University, Beijing 100084, China}

\author{Sang-Mok Cha}
\affil{Korea Astronomy and Space Science Institute, Daejon 34055,
Republic of Korea}
\affil{School of Space Research, Kyung Hee University, Yongin, Kyeonggi 
17104, Republic of Korea}

\author{Dong-Jin Kim}
\affil{Korea Astronomy and Space Science Institute, Daejon 34055,
Republic of Korea}

\author{Seung-Lee Kim}
\affil{Korea Astronomy and Space Science Institute, Daejon 34055,
Republic of Korea}
\affil{University of Science and Technology, Korea (UST) Gajeong-ro,
Yuseong-gu,  Daejeon 34113, Republic of Korea}

\author{Chung-Uk Lee}
\affil{Korea Astronomy and Space Science Institute, Daejon 34055,
Republic of Korea}

\author{Dong-Joo Lee}
\affil{Korea Astronomy and Space Science Institute, Daejon 34055,
Republic of Korea}

\author{Yongseok Lee}
\affil{Korea Astronomy and Space Science Institute, Daejon 34055,
Republic of Korea}
\affil{School of Space Research, Kyung Hee University, Yongin, Kyeonggi 
17104, Republic of Korea}

\author{Byeong-Gon Park}
\affil{Korea Astronomy and Space Science Institute, Daejon 34055,
Republic of Korea}
\affil{University of Science and Technology, Korea (UST) Gajeong-ro,
Yuseong-gu,  Daejeon 34113, Republic of Korea}

\author{Richard W. Pogge}
\affil{Department of Astronomy, Ohio State University, 140 W. 18th Ave., Columbus, OH 43210, USA}

\collaboration{(KMT Collaboration)}
\noaffiliation

\begin{abstract}
Some low-mass planets are expected to be ejected from their parent planetary systems during early stages of planetary system formation. According to planet formation theories, such as the core accretion theory, typical masses of ejected planets should be between $0.3$ and $1.0\,M_{\oplus}$. Although in practice such objects do not emit any light, they may be detected using gravitational microlensing via their light-bending gravity. Microlensing events due to terrestrial-mass rogue planets are expected to have extremely small angular Einstein radii ($\lesssim 1\,\mu\mathrm{as}$) and extremely short timescales ($\lesssim 0.1$\,day). Here, we present the discovery of the shortest-timescale microlensing event, OGLE-2016-BLG-1928, identified to date ($\tE\approx0.0288\,\mathrm{day}=41.5\,\mathrm{min}$). Thanks to the detection of finite-source effects in the light curve of the event, we were able to measure the angular Einstein radius of the lens $\thetaE=0.842\pm0.064$\,$\mu$as, making the event the most extreme short-timescale microlens discovered to date. Depending on its unknown distance, the lens may be a Mars- to Earth-mass object, with the former possibility favored by the \textit{Gaia} proper motion measurement of the source. The planet may be orbiting a star but we rule out the presence of stellar companions up to the projected distance of $\sim 8.0$\,au from the planet. Our discovery demonstrates that terrestrial-mass free-floating planets can be detected and characterized using microlensing. 
\end{abstract}


\section{Introduction} 
\label{sec:intro}

Thousands of extrasolar planets have been discovered to date. Although many of the known exoplanets do not resemble those in our solar system, they have one thing in common---they all orbit a star. However, theories of planet formation and evolution predict the existence of free-floating (rogue) planets, gravitationally unattached to any star. 

Exoplanets may be ejected from their parent planetary systems as a result of planet--planet scattering \citep{rasio1996,marzari1996,lin1997,chatterjee2008}. It is estimated that at least 75\% of systems with giant planets must have experienced planet--planet scattering in the past \citep[][and references therein]{raymond2020}. Dynamical interactions between giant planets inevitably lead to disruptions of orbits of inner smaller (rocky) planets \citep[e.g.,][]{veras2005,matsumura2013,carrera2016} and may lead to their ejection. In their population synthesis calculations (which are based on the core accretion theory of planet formation; \citealt{ida2013}), \citet{mao2016} found that typical masses of ejected planets are between $0.3$ and $1.0\,M_{\oplus}$. According to their model, rogue planets are more likely to form around massive stars, which are in turn more likely to host giant planets. A similar conclusion was reached by \citet{barclay2017} who carried out $N$-body simulations of terrestrial planet formation around solar-type stars. They found that in the presence of giant planets in such systems, a large fraction of the protoplanetary material is ejected, partly in the form of Mars-mass bodies ($\sim0.1-0.3\,M_{\oplus}$). Planets may also be liberated as a result of interactions in multiple-star systems \citep[e.g.,][]{kaib2013} and stellar clusters \citep[e.g.,][]{spurzem2009}, stellar flybys \citep[e.g.,][]{malmberg2011}, or the post-main-sequence evolution of the host star \citep[e.g.,][]{veras2011}.

Dark compact objects, such as rogue planets, may be in principle detected in gravitational microlensing events---microlensing does not depend on the brightness of a lensing object. However, typical Einstein timescales of microlensing events due to sub-Earth-mass objects are extremely short:
\begin{equation}
\tE=\frac{\thetaE}{\murel}=1.5\,\mathrm{hr}\left(\frac{M}{0.3\,M_{\oplus}}\right)^{1/2}\left(\frac{\pirel}{0.1\,\mathrm{mas}}\right)^{1/2}\left(\frac{\murel}{5\,\mathrm{mas\,yr}^{-1}}\right)^{-1}
\end{equation} 
rendering their detection difficult. (Here, $\thetaE$ is the angular Einstein radius, $\murel$ -- relative lens-source proper motion, $M$ -- mass of the lens, and $\pirel$ -- relative lens-source parallax.) If the radius of the source star is larger than the Einstein radius, the duration of microlensing events is extended thanks to finite-source effects \citep{gould1994,nemi1994,witt1994}. For sub-Earth-mass lenses, finite-source effects become important if the angular radius of the source, $\theta_*$, is of the order of the Einstein radius,
\begin{equation}
\thetaE=0.8\,\mu\mathrm{as}\left(\frac{M}{0.3\,M_{\oplus}}\right)^{1/2}\left(\frac{\pirel}{0.1\,\mathrm{mas}}\right)^{1/2}.
\end{equation} 
So far, only four short-timescale microlensing events exhibiting finite-source effects were identified \citep[i.e., OGLE-2012-BLG-1323, $\tE=0.155\pm0.005$\,day, $\thetaE=2.37\pm0.10$\,$\mu$as; OGLE-2016-BLG-1540, $\tE=0.320\pm0.003$\,day, $\thetaE=9.2\pm0.5$\,$\mu$as; OGLE-2019-BLG-0551, $\tE=0.381\pm0.017$\,day, $\thetaE=4.35\pm0.34$\,$\mu$as; KMT-2019-BLG-2073, $\tE=0.267\pm0.026$\,day, $\thetaE=4.77\pm0.19$\,$\mu$as;][]{mroz2018,mroz2019,mroz2020b,kim2020}. These events may be caused by unbound or wide-orbit ($\gtrsim 10$\,au) planets since microlensing observations alone are not able to rule out the presence of a distant stellar companion (as discussed in more detail by \citealt{mroz2020b}). These detections, together with short-timescale events found by \citet{mroz2017}, provide strong evidence for a large population of free-floating or wide-orbit planets in the Milky Way.

In this Letter, we present the discovery of the shortest-timescale microlensing event detected to date ($\tE=0.0288^{+0.0024}_{-0.0016}$\,day, $\thetaE=0.842\pm0.064$\,$\mu$as), which was likely caused by a Mars- to Earth-mass object.

\begin{figure}
\centering
\includegraphics[width=.48\textwidth]{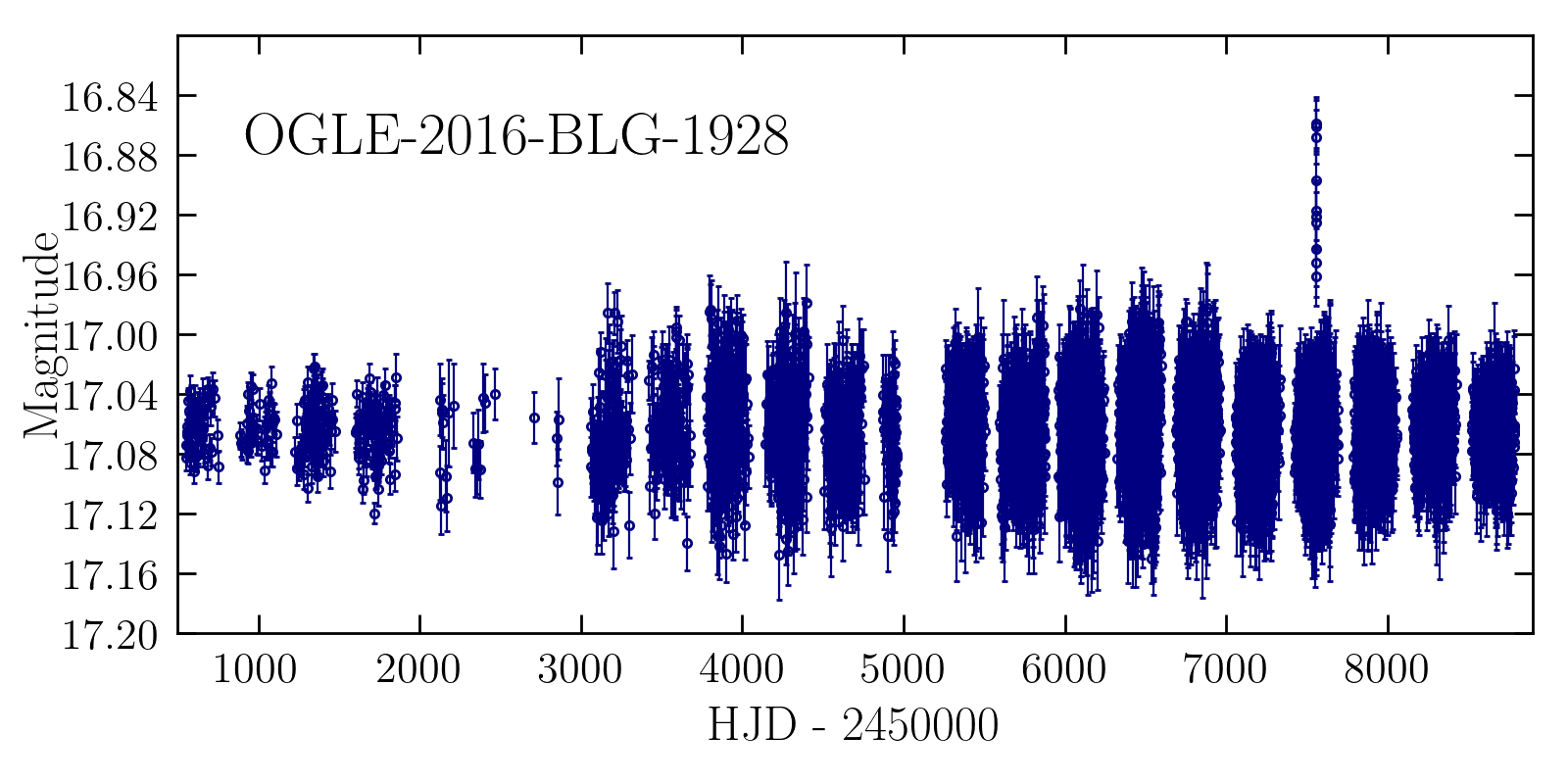}
\includegraphics[width=.48\textwidth]{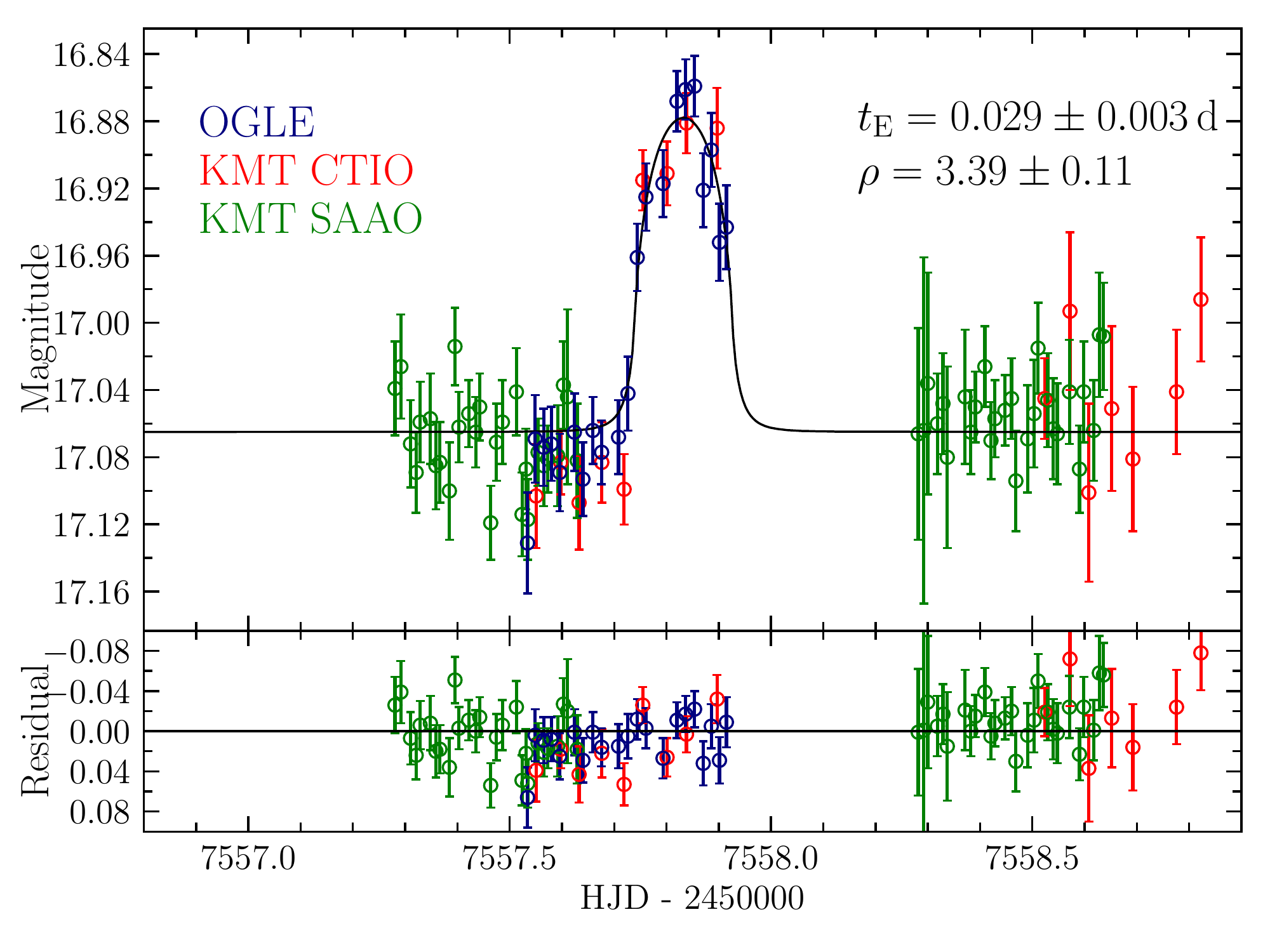}
\caption{Upper panel: 23~yr~long OGLE light curve of the microlensing event OGLE-2016-BLG-1928 reveals only one brightening that occurred on 2016~June~18 and lasted about 0.2~days. Lower panel: close-up of the magnified part of the light curve with the best-fitting microlensing model overplotted.}
\label{fig:model}
\end{figure}

\section{Data}

Microlensing event OGLE-2016-BLG-1928\footnote{This event was not detected in real-time by the OGLE Early Warning System \citep{udalski2003}. For consistency with previous works, we assigned it the name OGLE-2016-BLG-1928.} occurred on 2016~June~18 ($\mathrm{HJD}'=\mathrm{HJD}-2450000=7557.8$) on a bright star ($I=17.07$, $V-I=1.91$) located at the equatorial coordinates of RA = \ra{18}{01}{31}{25}, Decl. = \dec{-29}{07}{46}{2} (Galactic coordinates: $(l,b)=(1.596^{\circ},-3.094^{\circ})$). The event was found in data from the fourth phase of the Optical Gravitational Lensing Experiment \citep[OGLE;][]{udalski2015} as part of the search for wide-separation planetary systems (Poleski et al., in preparation) but it has been observed by OGLE since 1997. The event was located near the area that was extensively monitored during the Campaign~9 of the~\textit{K2} mission \citep{henderson2016} -- both by~$K2$ and numerous ground-based telescopes. However, only OGLE and one of the stations of the Korea Microlensing Telescope Network \citep[KMTNet;][]{kim2016} -- located in Cerro Tololo Interamerican Observatory (KMT~CTIO) -- captured the magnified part of the light curve (as shown in Figure~\ref{fig:model}). We used OGLE and KMT~CTIO data in single-lens models. For binary-lens modeling, we included additional data from the KMTNet telescope in the Southern African Astronomical Observatory (KMT~SAAO).
The event was also observed six times during $7557.87<\mathrm{HJD}'<7558.17$ by the Microlensing Observations in Astrophysics (MOA) survey \citep{bond2001} under adverse weather conditions (poor seeing and clouds), which prevented us from extracting useful data. All analyzed data were collected in the $I$-band. The photometry was extracted using custom implementations of the difference image analysis \citep{alard1998} method by \citet{wozniak2000} (OGLE) or \citet{albrow2009} (KMTNet).

\section{Single-lens Models}

The light curve of the event (Figure~\ref{fig:model}) can be well described by an extended-source point-lens model, which has four parameters: $t_0$ and $u_0$ -- time and projected separation during the closest approach between the lens and the center of the source, $\tE$ -- Einstein timescale, and $\rho=\theta_*/\thetaE$ -- which is the angular radius of the source $\theta_*$ expressed in $\thetaE$ units. The approximate values of these parameters can be estimated from the light curve without the need of sophisticated modeling \citep[cf.][]{mroz2020b}: the maximum magnification $A$ and duration $\Delta{t}$ of the event are related to $\rho\approx\sqrt{2/(A-1)}\approx3.1$ and $\tE\approx\Delta{t/2}\rho=0.03$\,days. Indeed, in our best-fitting model we measure $\rho=3.39^{+0.10}_{-0.11}$ and $\tE=0.0288^{+0.0024}_{-0.0016}$\,days (Table~\ref{tab:pars}). Microlensing magnifications are computed using the method described by \citet{bozza2018}, we assume a linear limb-darkening law with $\Gamma=0.46$ (as appropriate for the effective temperature of the source of $5000\pm200$\,K, see Section~\ref{sec:cmd}; \citealt{claret2011}). The best-fitting parameters and their uncertainties are estimated using the Markov Chain Monte Carlo sampler of \citet{foreman2013}.

During the modeling we fix the value of the dimensionless blending parameter $f_{\rm{s}}=1$, that is, we assume that the entire flux comes from the source star. Usually, for every data set, there are two additional parameters that describe the source flux $F_{\rm{s}}$ and unmagnified blend flux $F_{\rm{b}}$. We define $f_{\rm{s}}=F_{\rm{s}}/(F_{\rm{s}}+F_{\rm{b}})$. When both $F_{\rm{s}}$ and $F_{\rm{b}}$ were allowed to vary, the best-fitting solutions had large negative blending flux ($f_{\rm{s}}\gtrsim3$), such solutions are unphysical. The best-fitting model with $f_{\rm{s}}=1$ is disfavored by only $\Delta\chi^2=4.8$ which may be due to a statistical fluctuation or low-level systematics in the data. As demonstrated by \citet{mroz2020b}, blending does not influence the inferred value of $\thetaE$ provided that the blend and source have similar colors. Thus, in our final models, we kept $F_{\rm{b}}=0$ (that is, $f_{\rm{s}}=1$) constant, but we also added in quadrature 0.05~mag to the uncertainty of the source brightness. 
For comparison, the best-fit parameters for the free-blending fit (assuming $f_{\rm{s}}\leq1$) are also presented in Table~\ref{tab:pars}.
Blending may affect the characterization of the lens only if the blend and source have significantly different colors, as discussed in detail by \citet{mroz2020b} and \citet{kim2020}.

\begin{deluxetable*}{lcc|lc}
\tablecaption{Microlensing parameters for best-fit solutions\label{tab:pars}}
\tablehead{
\multicolumn{3}{c}{Point Lens} & \multicolumn{2}{c}{Binary Lens}\\
\colhead{Parameter} & \colhead{Value} & \colhead{Value} & \colhead{Parameter} & \colhead{Value}\\
\colhead{} & \colhead{($f_{\rm s} = 1$)} & \colhead{($f_{\rm s} \leq 1$)} & \colhead{} & \colhead{}
}
\startdata
$t_0$ (HJD$'$) & $7557.8332^{+0.0050}_{-0.0036}$ & $7557.8325\pm0.0037$ & $t_0~{\rm (HJD')}$ & $7593.31\pm0.74$ \\
$t_{\rm E}$ (days) & $0.0288^{+0.0024}_{-0.0016}$ & $0.0286 \pm 0.0020$ & $t_{\rm E}$ (days) & $1.93^{+0.13}_{-0.17}$ \\
$u_0$ & $0.59^{+0.58}_{-0.42}$ & $0.29^{+0.60}_{-0.29}$ & $u_0$ & $2.91^{+0.28}_{-0.39}$ \\
$\rho$ & $3.39^{+0.10}_{-0.11}$ & $3.31^{+0.12}_{-0.25}$ & $\rho$ & $0.0518\pm0.0048$ \\
$I_{\rm s}$ & $17.07 \pm 0.05$ & $17.11^{+0.16}_{-0.05}$ & $s$ & $18.7^{+1.7}_{-1.1}$ \\
$f_{\rm s}$  & $1.0$ (fixed) & $0.96^{+0.05}_{-0.13}$ & $q$ & $0.34^{+0.17}_{-0.10} \times10^{-3}$ \\
   & \phn  & \phn & $\alpha$ (deg) & $188.8\pm1.2$
\enddata
\tablecomments{HJD$'$=HJD-2450000. $f_{\rm s}=F_{\rm{s}}/(F_{\rm{s}}+F_{\rm{b}})$ is the dimensionless blending parameter.}
\end{deluxetable*}

\section{Binary-lens Models}

The light curve of OGLE-2016-BLG-1928 shows a clear signal from a low-mass planet, but it does not show an obvious signal from a host of the planet. To search for a host, we fitted a binary-lens model to the data. The binary-lens model has three parameters more than the single-lens model and these are: $s$ -- the projected separation between the planet and host expressed in Einstein radii (of the total mass of the system), $q$ -- planet to host mass ratio, and $\alpha$ -- angle between the binary axis and the source trajectory. We started the search for binary-lens models by defining $t_0$, $u_0$, and $t_{\rm E}$ relative to the planet \citep{han2006,mroz2020b}, because then the values of the four parameters $(t_0,u_0,\tE,\rho)$ are well constrained by the light curve. In order to speed-up calculations, we neglected limb-darkening of the source. The magnification of the finite-source binary-lens model was evaluated using the method presented by \citet{bozza2010} and \citet{bozza2018}. The fitting was done using the MulensModel code by \citet{poleski_yee2019}. After these initial fits converged, we reran the fits in standard parameterization ($t_0$, $u_0$ defined relative to the center of mass, and $t_{\rm E}$ relative to the total mass of the system) and including limb-darkening of the source. 

\begin{figure}
\centering
\includegraphics[width=.5\textwidth]{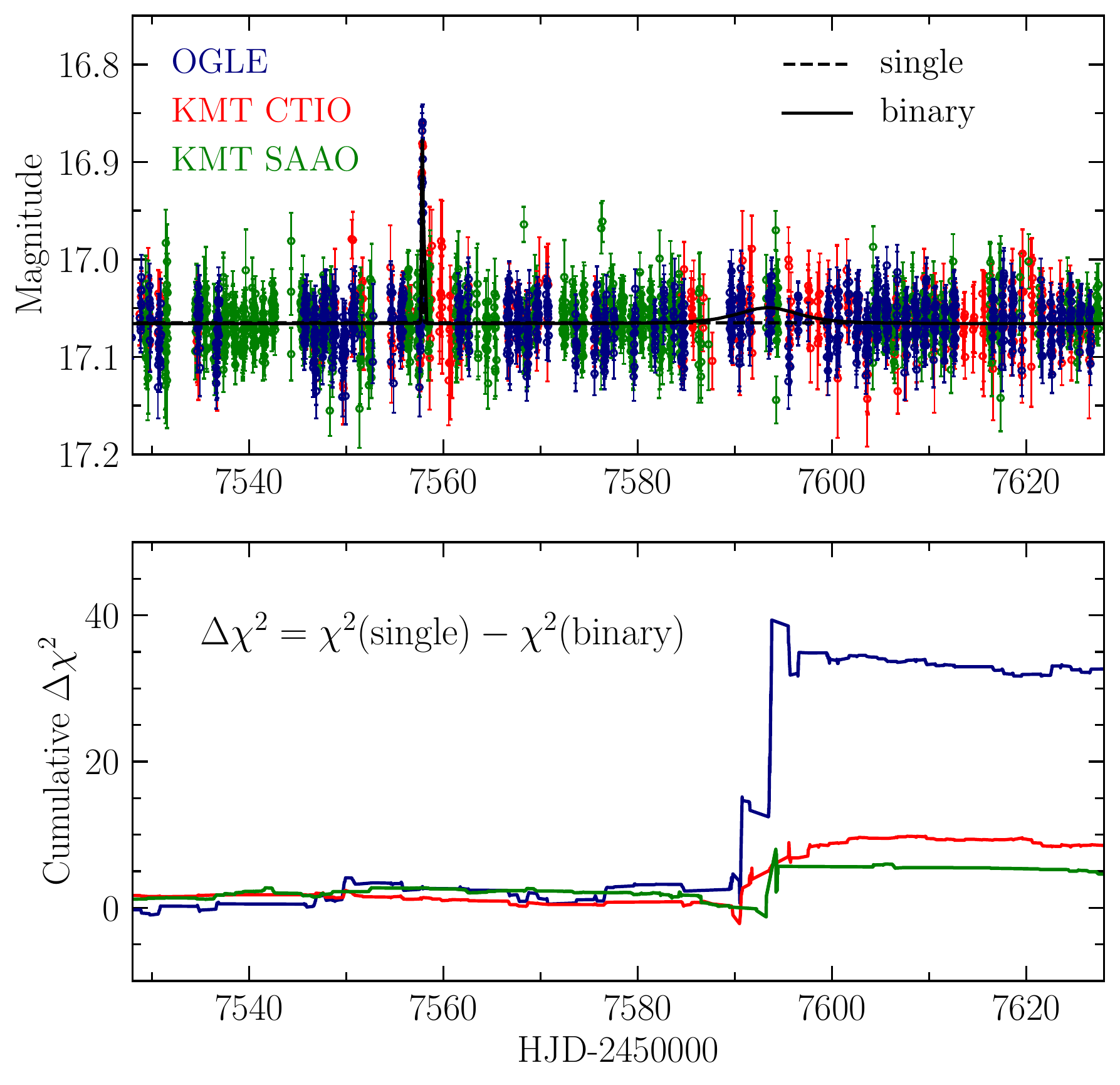}
\caption{Upper panel: comparison between the single- (dashed line) and binary-lens (solid line) models. Lower panel: cumulative distribution of $\Delta\chi^2$ between these models.}
\label{fig:binary}
\end{figure}

The parameters of the best-fit binary-lens model are presented in Table~\ref{tab:pars}. When compared to the single-lens model, the $\chi^2$ improves by 44.2. The main difference between single-lens and binary-lens models is the presence of a low-amplitude bump at ${\rm HJD'}\approx7593$. The $\chi^2$ improvement, however, comes mostly from one observatory (OGLE) from one night and the KMT data from that night do not corroborate the signal (Figure~\ref{fig:binary}). The bump could have an origin other than the microlensing of planet host: it could be produced by low-level fluctuations in the light curve (either of instrumental origin, intrinsic variability, or a combination of both). We cannot judge the reliability of the binary-lens fit using the Bayesian approach because we cannot present a meaningful prior on such a binary-lens model. Instead, we decided to check $\Delta\chi^2$ relative to a constant brightness model and determine whether similar bumps are present in other seasons of the OGLE-IV data. We fitted point-source point-lens models to each of the observing seasons 2010--2015 and 2017--2019. For each season ($i$), we calculated $\chi^2$ difference between the above fit and a model of constant brightness ($\Delta\chi^2_i$). In order to compare these values with the bump in 2016 we normalized them by the number of epochs in the given season: $\Delta\chi^2_i N_{2016}/N_i$, where $N_{2016}=1621$. The results of these calculations are presented in Table~\ref{tab:trends}. The $\Delta\chi^2_{2016}$ between point-lens and binary-lens models for 2016 data is 32.5 when only OGLE data are considered. We see that in four out of nine other seasons the $\Delta\chi^2_i N_{2016}/N_i$ is higher than 32.5, hence, the probability that the bump detected in binary-lens analysis is just the manifestation of low-level fluctuations as seen in other seasons is $4/9=44\%$.

Moreover, the parameters of the binary-lens model point to a very unusual system, suggesting that the bump in the light curve is not due to microlensing. The Einstein timescale of $1.9~{\rm days}$ is extremely short and suggests that the host has a mass of a few Jupiter masses, and hence may be a planet as well. Additionally, the projected separation $s\approx19$ is almost four times larger than the widest separation microlensing planet currently known (OGLE-2008-BLG-092, $s=5.3$; \citealt{poleski2014Uran}).

Since there is no strong evidence for a host star from the microlensing light curve, we use the method of \citet{mroz2018,mroz2019} to estimate lower limits on the projected star--planet separation of a putative host star. We consider a $0.3\,M_{\odot}$ host located either in the Galactic disk ($\pi_{\rm rel}=0.1\,$mas) or in the bulge ($\pi_{\rm rel}=0.016$\,mas), which correspond to $\theta_{\rm E,host}=0.49$\,mas or 0.20\,mas, respectively. Then, we simulate synthetic OGLE light curves (spanning from 2010~March~5 through 2019~October~30) assuming $q=(\thetaE/\theta_{\rm E,host})^2$, $1\leq{s}\leq10$, and $0\leq\alpha\leq2\pi$. For each pair of $(q,s)$ we calculate the fraction of light curves that show signatures of the host star. We find a 90\%~lower limit on the projected host separation of 3.6~Einstein radii for the lens located in the disk ($\pi_{\rm rel}=0.1\,$mas) and 3.3~Einstein radii for the bulge lens ($\pi_{\rm rel}=0.016\,$mas). These limits translate to 8.0\,au and 4.6\,au, respectively.

We also searched for binary-lens solutions in which the observed brightening is due to a~cusp crossing. We did a~grid search with $-2\leq\log{q}\leq0$ and $0.2\leq{s}\leq3$. We considered only trajectories that cross the caustic twice during the night of $\textrm{HJD}'=7557$ and parameterized the models using the \citet{cassan2008} approach. In the best-fitting model ($\tE=0.020$\,day,$\rho=2.4$,$s=4.4$,$q=0.86$), the source envelops the caustics. Although this model is better by $\Delta\chi^2=10.0$ than the best single-lens model, we find it unlikely (this is essentially the same model but with an~extra body). We found $\tE<0.05$\,days for all binary-lens models in the grid search.

\begin{table}
\centering
\caption{Statistics of trends seen in OGLE data in seasons 2010--2015 and 2017--2019.}
\label{tab:trends}
\begin{tabular}{rrr}
\hline
year & $N_i$ & $\Delta\chi^2_i N_{2016} / N_i$ \\
\hline
2010 &  651 & 45.2 \\
2011 &  813 & 26.2 \\
2012 & 2317 & 51.7 \\
2013 & 2309 & 41.7 \\
2014 & 2277 & 13.9 \\
2015 & 2170 & 10.2 \\
2017 &  707 & 68.1 \\
2018 &  724 & 30.8 \\
2019 &  824 & 16.7 \\
\hline
\end{tabular}
\end{table}

\begin{deluxetable}{lc}
\tablecaption{Physical parameters for the source and lens for point lens ($f_{\rm s}=1.0$ solution)\label{tab:physical}}
\tablehead{
\colhead{Parameter} & \colhead{Value}}
\startdata
Source: & \\
$I_{\rm s,0}$      & $15.78 \pm 0.08$ \\
$(V-I)_{\rm s,0}$  & $0.93 \pm 0.02$ \\
$(V-K)_{\rm s,0}$  & $2.12 \pm 0.07$ \\
$T_{\rm eff}$ (K)  & $5000 \pm 200$ \\
$\Gamma$ (limb-darkening, $I$ band) & 0.46 \\
$\theta_*$ ($\mu$as) & $2.85 \pm 0.20$ \\
$\mu_{l}$ (mas\,yr$^{-1}$) & $-6.12 \pm 1.03$ \\
$\mu_{b}$ (mas\,yr$^{-1}$) & $-0.13 \pm 0.81$ \\
\hline
Lens: & \\
$\theta_{\rm E}$ ($\mu$as)       & $0.842 \pm 0.064$ \\
$\mu_{\rm rel}$ (mas\,yr$^{-1}$) & $10.6 \pm 1.0$ \\
\enddata
\end{deluxetable}

\section{Physical Parameters}
\label{sec:cmd}

\begin{figure}
\centering
\includegraphics[width=.5\textwidth]{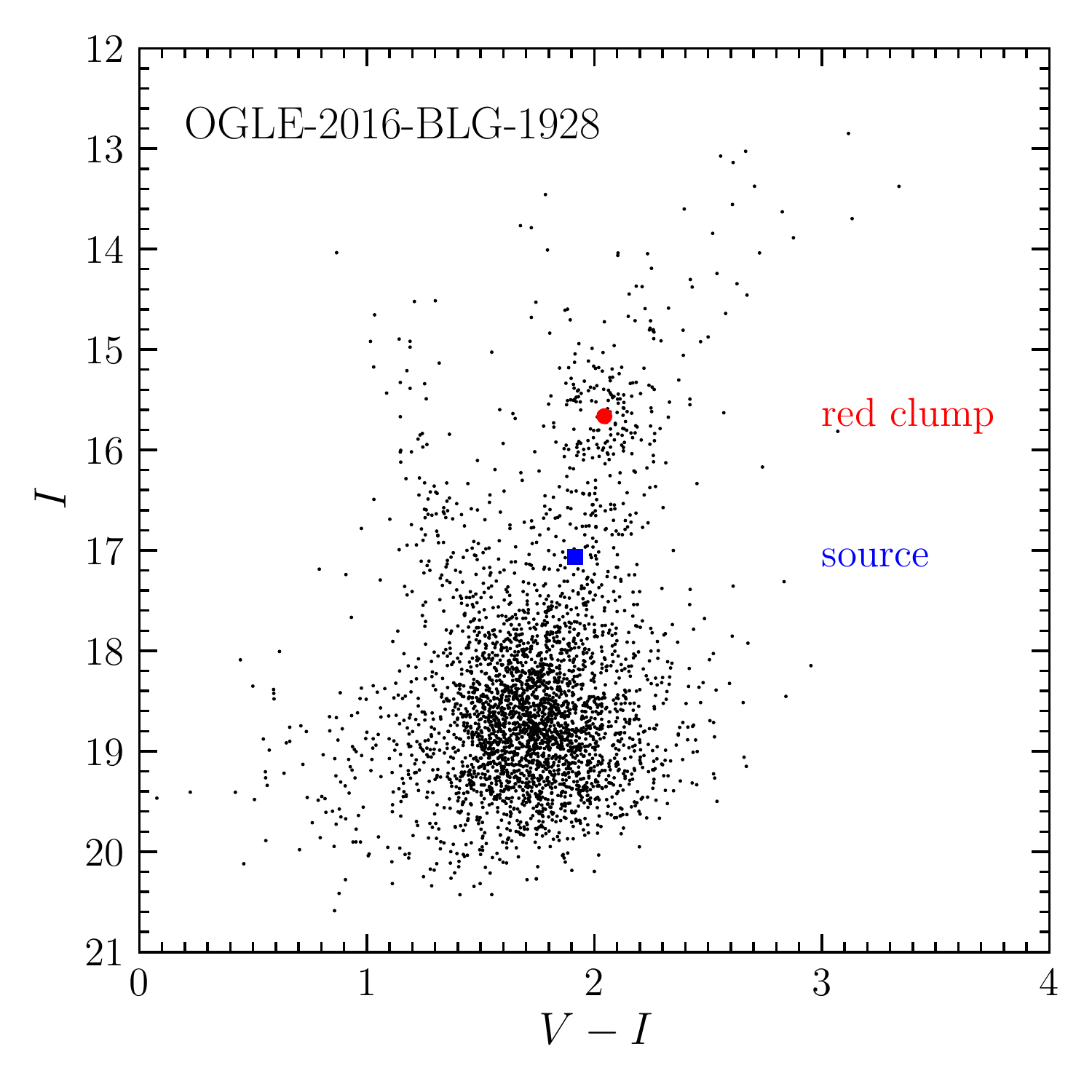}
\caption{Color--magnitude diagram of stars located within $2'\times 2'$ of the microlensing event OGLE-2016-BLG-1928.}
\label{fig:cmd}
\end{figure}

The light curve of the event OGLE-2016-BLG-1928 (Figure~\ref{fig:model}) exhibits prominent finite-source effects that enable us to measure the angular Einstein radius of the lens provided that the angular radius of the source star is known: $\thetaE=\theta_*/\rho$. We use the color--surface brightness relation of \citet{pietrzynski2019} to calculate $\theta_*$. To determine the dereddened color $(V-I)_{\mathrm{s},0}$ and brightness $I_{\mathrm{s},0}$ of the source, we use the standard method of \citet{yoo2004}. We measure that the source is $\Delta(V-I)=-0.13\pm0.02$ bluer and $\Delta I=1.40\pm0.09$ fainter than the red clump centroid in the color--magnitude diagram (Figure~\ref{fig:cmd}). Because the dereddened color ($(V-I)_{\mathrm{RC},0}=1.06$) and brightness ($I_{\mathrm{RC},0}=14.38$) of red clump stars in this direction are known \citep{bensby2011,nataf2013}, we measure $(V-I)_{\mathrm{s},0}=0.93\pm0.02$ and $I_{\mathrm{s},0}=15.78\pm0.09$. Subsequently, we determine $(V-K)_{\mathrm{s},0}=2.12\pm0.07$ using the  color--color relations of \citet{bessell1988} and $\theta_*=2.85\pm0.20$\,$\mu$as using the color--surface brightness relation of \citet{pietrzynski2019}.

\begin{figure}
\centering
\includegraphics[width=.5\textwidth]{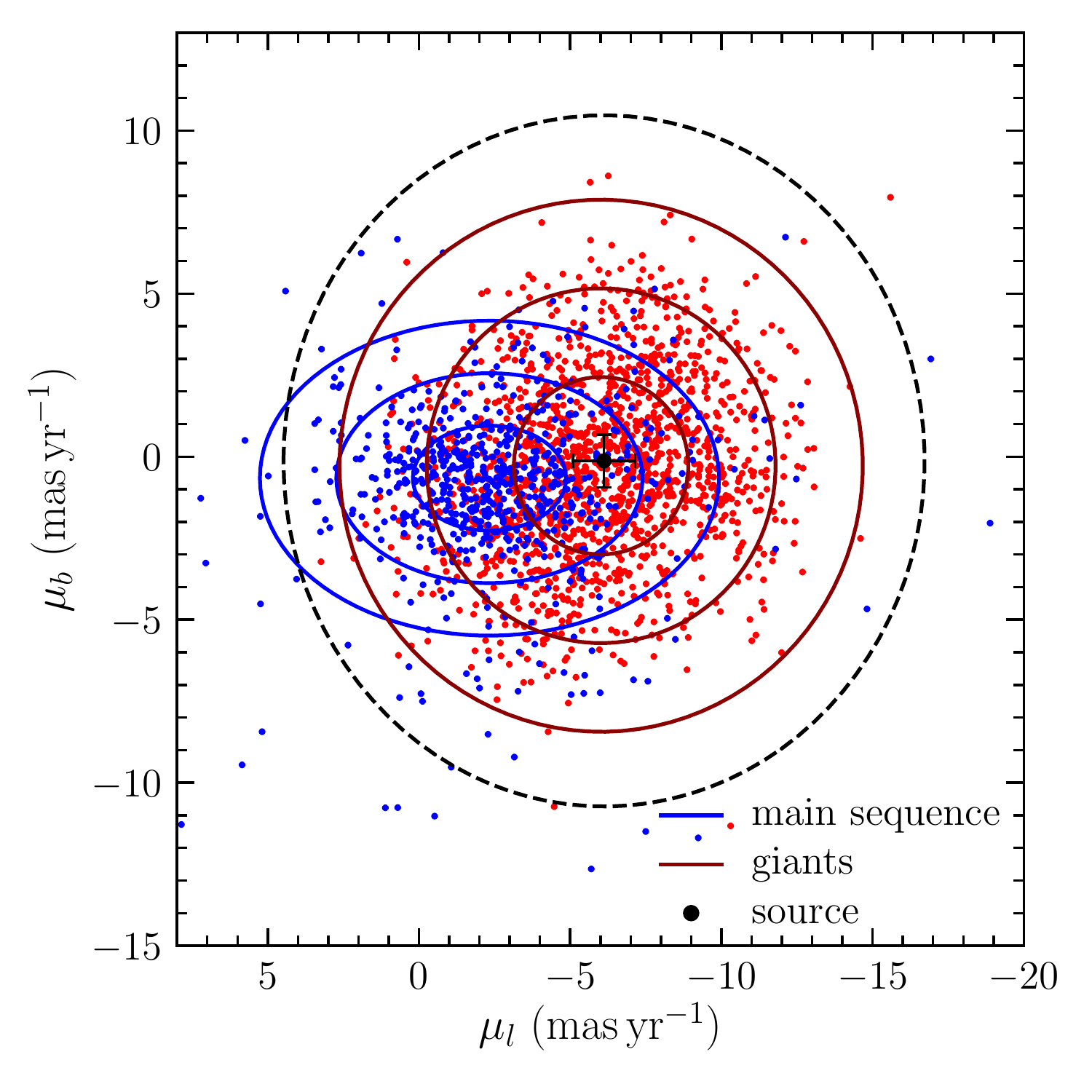}
\caption{\textit{Gaia} proper motions of stars located within $5'$ of the event \citep{gaia2018}. Red giant stars (representing the Galactic bulge population) are marked by red dots, while main-sequence stars (Galactic disk population) are marked by blue dots. Solid contours mark $(1,2,3)\sigma$ error ellipses based on the scatter in the distribution. The proper motion of the source and its position on the color--magnitude diagram are consistent with those of Galactic bulge stars. The relative lens-source proper motion is $\murel=10.6\pm1.0\,\mathrm{mas\,yr}^{-1}$, so the lens should be located on a dashed circle.}
\label{fig:pm}
\end{figure}

The calculation above is based on two assumptions. First, we assume that the source star and red clump stars are reddened by the same amount. Because the \textit{Gaia} proper motion of the source relative to the mean proper motion of red clump stars is only $0.18\,\mathrm{mas\,yr}^{-1}$ \citep{gaia2018} (Figure~\ref{fig:pm}), the source is likely located in the Galactic bulge and so the first assumption holds true. Second, we assume that the color of the source is equal to that of the baseline object. Neither OGLE nor KMTNet observed the magnified part of the event in the $V$-band filter, which prevents us from directly calculating the color of the source. Because the best-fitting model evinces no evidence for blended light from unresolved ambient stars, our best estimate of the color of the source is the color of the baseline object.

Having the angular radius of the source star measured, we can calculate the angular Einstein radius:
\begin{equation}
\thetaE=\frac{\theta_*}{\rho}=0.842\pm0.064\,\mu\mathrm{as}
\end{equation}
and the relative lens-source proper motion (in the geocentric frame):
\begin{equation}
\murel=\frac{\thetaE}{\tE}=10.6\pm1.0\,\mathrm{mas\,yr}^{-1}.
\end{equation}

\section{Discussion}

With the Einstein timescale of $\tE=0.0288^{+0.0024}_{-0.0016}\,\mathrm{d}=41.5^{+3.5}_{-2.3}\,\mathrm{min}$ and the angular Einstein radius of $\thetaE=0.842\pm0.064\,\mu\mathrm{as}$, OGLE-2016-BLG-1928 is the most extreme short-timescale microlensing event discovered to date. The mass of the lens cannot be determined because the relative lens-source parallax cannot be measured:
\begin{equation}
M = \frac{\thetaE^2}{\kappa\pirel},
\end{equation}
where $\kappa=8.144\,\mathrm{mas}\,\mathrm{M_{\odot}}^{-1}$. If the lens is located in the Galactic disk ($\pirel\approx0.1\,\mathrm{mas}$), then $M\approx0.3\,M_{\oplus}$ (which is approximately three Mars masses). The lens located in the Galactic bulge (typically $\pirel\approx0.016\,\mathrm{mas}$) would be more massive ($M\approx2\,M_{\oplus}$). 

The \textit{Gaia} proper motion of the source (Table~\ref{tab:physical}) favors the interpretation that the lens is located in the Galactic disk, so the lens should be a~sub-Earth-mass object. The proper motion of the source is consistent with that of red clump stars (Figure~\ref{fig:pm}), and the relative lens-source proper motion is $\murel=10.6\pm1.0\,\mathrm{mas\,yr}^{-1}$. In order to have this relative proper motion, the lens should lie in the region around a dashed circle marked in Figure~\ref{fig:pm} and there are virtually no Galactic bulge stars in this region (while there exist some disk stars).
To quantify this, we can directly measure the proper motion distribution of Galactic bulge stars using \textit{Gaia} data (Figure~\ref{fig:pm}). This distribution can be approximated as a Gaussian with dispersions of $2.884\pm0.052$ and $2.720\pm0.049\,\mathrm{mas\,yr}^{-1}$ in the $l$ and $b$ directions, respectively. Thus the probability that the proper motion of the lens is consistent with that of bulge stars is smaller than $2\times10^{-4}$.

The lens in OGLE-2016-BLG-1928 is likely a sub-Earth-mass object, one of the lowest-mass objects ever found by microlensing. As in the case of other short-timescale microlensing events \citep{sumi2011,mroz2018,mroz2019,mroz2020b,kim2020}, we cannot rule out the presence of a distant stellar companion. We conducted an extensive search for possible binary-lens models -- we found that the best-fitting binary-lens model is preferred by $\Delta\chi^2=44.2$ over the single-lens model. Although this appears to be statistically significant, we found that the source exhibits low-level fluctuations, which may mimic a microlensing signal from a host star. Thus, we do not find any significant evidence for the host star up to the projected distance of 8.0\,au from the planet (assuming $\pirel=0.1\,$mas).

The properties of OGLE-2016-BLG-1928 place it at the edge of current limits of detecting short-timescale microlensing events and highlight the challenges that will be faced by future surveys for extremely short-timescale events (for example, by the \textit{Nancy Grace Roman Space Telescope}, formerly known as \textit{WFIRST}, \citealt{johnson2020}). Despite the fact that the event was located in high-cadence survey fields, only 15 data points were magnified (11 from OGLE and 4 from KMTNet), rendering the event difficult to detect. In particular, the declining part of the light curve is not fully covered with observations (Figure~\ref{fig:model}).

This raises the question of whether the observed light curve is due to a~genuine microlensing event in the first place. The source star is located in the red giant branch in the color--magnitude diagram (Figure~\ref{fig:cmd}), and some giants are known to produce stellar flares \citep[e.g.,][]{vd2017,iwanek2019}. However, the properties of the event (its duration, amplitude, and light-curve shape) do not match those of flaring stars. For example, \citet{balona2015} compiled an~atlas of stellar flares observed by the \textit{Kepler} satellite in short-cadence mode. They found that 97.8\%~of stellar flares are shorter than 0.2~day and that the light-curve shapes and amplitudes of the~remaining 2.2\% do not match those of OGLE-2016-BLG-1928. We also note that the event has been observed by OGLE since~1997 and there is no evidence for other flares (nor periodic variability due to star spots) in the archival data, suggesting the object is unlikely to be a flaring star.

Another issue resulting from the short duration of the event is the lack of color measurements while the source is magnified. According to \citet{mroz2020b}, in microlensing events exhibiting strong finite-source effects, the angular Einstein radius depends on the surface brightness of the source, which make color measurements critical for determining $\thetaE$. In the present case, we assumed that the source color is equal to the color of the baseline object, which is motivated by the lack of evidence for the blended light in the best-fitting models. This issue may become more important for the \textit{Roman} telescope which is being designed to carry out observations in~\textit{W146} filter with a 15~minute cadence and in~\textit{Z087} filter with a 12~hr cadence. \citet{johnson2020} estimate that only approximately 10\% of short-timescale events due to $1\,M_{\oplus}$ lenses would have a color measurement. We thus advocate that the frequency of \textit{Z087} filter observations should be increased. 

The discovery of OGLE-2016-BLG-1928 demonstrates that current microlensing surveys are capable of finding extremely -short-timescale events. Although the mass of the lens cannot be unambiguously measured, properties of the event are consistent with the lens being a sub-Earth-mass object with no stellar companion up to the projected distance of $\sim8$\,au from the planet. Thus, the lens is one of the best candidates for a terrestrial-mass rogue planet detected to date. This population of low-mass free-floating (or wide-orbit) planets may be further explored by the upcoming microlensing experiments.

\section*{Acknowledgements}
The OGLE project has received funding from the National Science Centre, Poland, grant MAESTRO 2014/14/A/ST9/00121 to A.U. R.P. was supported by the Polish National Agency for Academic Exchange via Polish Returns 2019 grant. Work by A.G. was supported by JPL grant 1500811. Work by C.H. was supported by the grants of National Research Foundation of Korea (2017R1A4A1015178 and 2020R1A4A2002885).
This research has made use of the KMTNet system operated by the Korea Astronomy and Space Science Institute (KASI) and the data were obtained at three host sites of CTIO in Chile, SAAO in South Africa, and SSO in Australia.

\bibliographystyle{aasjournal}

\end{document}